\documentclass[12pt]{article}
\newcommand{\Fc}{\mathcal{F}}
\newcommand{\AP}{\alpha^{\prime}}
\newcommand{\pd}{\partial}

\usepackage{amssymb,amsmath}

\begin{document}
\title{\textbf{Gravitational models with non-local scalar fields}}


\author{Sergey Yu. Vernov\\
        Skobeltsyn Institute of Nuclear
    Physics, Moscow State University,\\ Leninskie Gory 1, 119991, Moscow,
    Russia\\
        \small E-mail: svernov@theory.sinp.msu.ru}

\date{ }

\maketitle

\begin{abstract}
 A general class of $f(R)$ gravity models with minimally
coupling a nonlocal scalar field is considered. The
  Ostrogradski representation for nonlocal gravitational models with a quadratic potential and
  the way of its localization are proposed. We study the action with
  an arbitrary analytic function $\Fc(\Box_g)$, which has both simple
  and double roots. The way of localization allows to find particular solutions
  of nonlocal equations of gravity.
\end{abstract}

\section{Introduction}

Recently a new class of cosmological models based on the string field
theory (SFT)~\cite{review-sft} and the $p$-adic string theory emerges
and attracts a lot of attention \cite{IA1}--\cite{GK}. It is known that
the SFT and the $p$-adic string theory are UV-complete ones. Thus, one
can expect that resulting (effective) models should be free of
pathologies. These models exhibit one general non-standard property,
namely, their actions have terms with infinitely many derivatives, i.e.
nonlocal terms. The higher derivative terms usually produce phantom
fields \cite{Ostrogradski:1850,PaisU} (see also~\cite{AV-NEC}).  Models
that includes phantoms violate the null energy condition (NEC), and,
therefore, are unstable. Models with higher derivative terms produce
also well-known problems with quantum instability~\cite{AV-NEC}.

To obtain a stable model with the NEC violation (the state parameter
$w_{\mathrm{DE}}<-1$) one should construct this model as an effective
model, connected with the fundamental theory, which is stable and
admits quantization. With the lack of quantum gravity, we can just
trust string theory or deal with an effective theory admitting the UV
completion.

The purpose of this paper is to study $f(R)$ gravity models with a
nonlocal scalar field. We consider a general form of nonlocal action
for the scalar field with a quadratic potential, keeping the main
ingredient, the analytic function $\Fc(\Box_g)$, which in fact produces
the nonlocality, almost unrestricted.


\section{Nonlocal gravitation models}

The SFT inspired nonlocal gravitation models~\cite{IA1} are introduced
as a sum of the SFT action of the tachyon field $\phi$ plus the gravity
part of the action. One cannot deduce this form of the action from the
SFT. In this paper we study the $f(R)$ gravity, which is a
straightforward modification of the general relativity. We consider the
following action:
\begin{equation}
S_f=\int d^4x \sqrt{-g}\left(\frac{f(L^2R)}{16\pi
G_NL^2}+\frac{1}{\alpha^{\prime}g_o^2}\left(\frac{1}{2}\phi\,\Fc\left(\alpha^{\prime}\Box_g\right)\phi
-V(\phi) \right)-\Lambda\right),
\end{equation}
where $f(L^2R)$ is an arbitrary differentiable function. We use the
signature $(-,+,+,+)$, $g_{\mu\nu}$ is the metric tensor, $G_N$ is the
Newtonian constant. The potential $V(\phi)$ is a quadratic polynomial
$V(\phi)=C_2\phi^2+C_1\phi+C_0$, where $C_2$, $C_1$, and $C_0$ are
arbitrary real constants.

The function $\Fc$ is assumed to be analytic at all finite points of
the complex plane, in other words, to be an entire function. The
function $\Fc$ can be represented by the convergent series expansion:
$\Fc(\Box_g)=\sum\limits_{n=0}^{\infty}f_n\Box_g^{\;n}$. The
Weierstrass factorization theorem asserts that the function $\Fc$ can
be represented by a product involving its zeroes $J_k$:
\begin{equation}
\Fc(J)=J^me^{Y(J)}\prod_{k=1}^\infty\left(1-\frac{J}{J_k}\right)e^{\frac{J}{J_k}+\frac{J^2}{2J_k^2}
+\dots+\frac{1}{p_k}\left(\frac{J}{J_k}\right)^{p_k}},
\end{equation}
where $m$ is an order of the root $J=0$ ($m$ can be equal to zero),
$Y(J)$ is an entire function, natural numbers $p_n$ are chosen such
that the series
$\sum\limits_{n=1}^\infty\left(\frac{J}{J_n}\right)^{p_n+1}$
 is an absolutely and uniformly convergent one.

Scalar fields $\phi$ (associated with the open string tachyon) is
dimensionless, while $[\AP]=\mbox{length}^2$, $[L]=\mbox{length}$ and
$[g_o]=\mbox{length}$. Let us introduce dimensionless coordinates
$\bar{x}_\mu=x_\mu/\sqrt{\alpha'}$, the dimensionless Newtonian
constant $\bar{G}_N=G_N/\alpha'$, the dimensionless parameter $\bar
L=L/\sqrt{\alpha'}$, and the dimensionless open string coupling
constant $\bar g_o=g_o/\sqrt{\AP}$. The dimensionless cosmological
constant $\bar\Lambda=\Lambda{\AP}^2$, $\bar{R}$ is the curvature
scalar in the coordinates $\bar{x}_\mu$:
\begin{equation}
S_f=\int d^4 \bar{x} \sqrt{-g}\left(\frac{f(\bar{L}^2 \bar{R})}{16\pi
\bar{G}_N\bar{L}^2}+\frac{1}{\bar{g}_o^2}\left(\frac{1}{2}\phi\,\Fc\left(\bar{\Box}_g\right)\phi
-V(\phi) \right)-\bar{\Lambda}\right). \label{action_model2}
\end{equation}
In the following formulae we omit bars, but use only dimensionless
coordinates and parameters.

It is well-known~\cite{Mukhanov1} that at $f'(R)>0$ any $f(R)$ gravity
models in the metric variational approach are equivalent to the
Einstein gravity with a scalar field\footnote{There are two types of
$f(R)$ gravity: the metric variational approach and the Palatini
formalism. In the first case the equations of motion are obtained by
variation with respect to metric. Connections are the function of
metric in this formalism. In the Palatini formalism one should vary the
action independently with respect to metric and the connections.}. In
the metric variational approach the equations of gravity are as
follows:
\begin{equation}
\label{fr_equ} G_{\mu\nu}\equiv f'(R)R_{\mu\nu}-
\frac{f(R)}{2}g_{\mu\nu}-D_\mu
\partial_\nu f'(R)+g_{\mu\nu}\Box_g f'(R)=8\pi
G_N T_{\mu\nu}, \quad
  \Fc(\Box_g)\phi=\frac{dV}{d\phi},
\end{equation}
where the energy--momentum (stress) tensor $T_{\mu\nu}$ is:
\begin{equation}
  \label{TEV}
  T_{\mu\nu}\equiv{}-\frac{2}{\sqrt{-g}}\frac{\delta{S}}{\delta
    g^{\mu\nu}}
  =\frac{1}{g_o^2}\Bigl(E_{\mu\nu}+E_{\nu\mu}-g_{\mu\nu}\left(g^{\rho\sigma}
    E_{\rho\sigma}+W\right)\Bigr),
\end{equation}
\begin{equation}
  E_{\mu\nu}\equiv\frac{1}{2}\sum_{n=1}^\infty
  f_n\sum_{l=0}^{n-1}\partial_\mu\Box_g^l\phi\partial_\nu\Box_g^{n-1-l}\phi,\quad
  W\equiv\frac{1}{2}\sum_{n=2}^\infty
  f_n\sum_{l=1}^{n-1}\Box_g^l\phi\Box_g^{n-l}\phi-\frac{f_0}{2}\phi^2+C_1\phi.
\end{equation}

\section{Localization of nonlocal gravitational actions}

The Ostrogradski representation has been proposed for polynomial
$\Fc(\Box)$ in the Minkowski space-time~\cite{Ostrogradski:1850,PaisU}.
Our goal is to generalize this result on gravitational models with an
arbitrary analytic function $\Fc(\Box)$ with simple and double roots.
We also generalize the Ostrogradski representation on the models with a
linear potential. The nonlocal cosmological models with quadratic
potentials have been studied
in~\cite{Koshelev07,AJV0701,AJV0711,MN,KV,Vernov2010,VernovSQS}.

Let us start with the case $C_1=0$. We consider a function $\Fc(J)$,
which has simple roots $J_i$ and double roots $\tilde{J}_k$, and the
function
\begin{equation}
  \label{phi0}
  \phi_0=\sum\limits_{i=1}^{N_1}\phi_i+\sum\limits_{k=1}^{N_2}\tilde\phi_k,
\end{equation}
where
\begin{equation}
  (\Box_g-J_i)\phi_i=0 \quad\mbox{and}\quad (\Box_g-\tilde{J}_k)^2\tilde\phi_k=0\quad\Leftrightarrow\quad
(\Box_g-\tilde{J_k})\tilde\phi_k=\varphi_k,\quad
  (\Box_g-\tilde{J_k})\varphi_k=0.
  \label{equphi}
\end{equation}
Without loss of generality we assume that for any $i_1$ and $i_2\neq
i_1$ conditions $J_{i_1}\neq J_{i_2}$ and
${\tilde{J}}_{i_1}\neq{\tilde{J}}_{i_2}$ are satisfied.

The energy--momentum tensor, which corresponds to $\phi_0$, has the
following form:
\begin{equation}
  T_{\mu\nu}\left(\phi_0\right)=
  T_{\mu\nu}\left(\sum\limits_{i=1}^{N_1}\phi_i+\sum\limits_{k=1}^{N_2}\tilde\phi_k\right)=
  \sum\limits_{i=1}^{N_1}T_{\mu\nu}(\phi_i)+\sum\limits_{k=1}^{N_2}T_{\mu\nu}(\tilde\phi_k),
  \label{Tmunugen}
\end{equation}
where all $T_{\mu\nu}$ are given by (\ref{TEV}) and
\begin{equation}
  E_{\mu\nu}(\phi_i)=\frac{{
      \Fc'(J_i)}}{2}\partial_{\mu}\phi_i\partial_{\nu}\phi_i,\quad
  E_{\mu\nu}(\tilde\phi_k)= \frac{{
      \Fc''(\tilde{J}_k)}}{4}\left(\partial_\mu\tilde\phi_k\partial_\nu\varphi_k
      +\partial_\nu\tilde\phi_k\partial_\mu\varphi_k\right)+
  \frac{\Fc'''(\tilde{J}_k)}{12}\partial_\mu\varphi_k\partial_\nu\varphi_k,
\end{equation}
\begin{equation}
  \label{Vdr} W(\phi_i)=\frac{J_i \Fc'(J_i)}{2}\phi_i^2,\quad W(\tilde{\phi}_k)=\frac{\tilde{J}_k
    \Fc''(\tilde{J}_k)}{2}\tilde\phi_k\varphi_k+ \left(\frac{{\tilde{J}_k
        \Fc'''(\tilde{J}_k)}}{12}+\frac{{
        \Fc''(\tilde{J}_k)}}{4}\right)\varphi_k^2,
\end{equation}
where a prime denotes a derivative with respect to $J$: $\Fc'\equiv
\frac{d\Fc}{dJ}$, \ $\Fc''\equiv \frac{d^2\Fc}{dJ^2}$ and $\Fc'''\equiv
\frac{d^3 \Fc}{dJ^3}$.

Considering the following local action
\begin{equation}
  S_{loc}=\int d^4x\sqrt{-g}\left(\frac{f(R)}{16\pi
      G_N}-\Lambda\right)+\sum_{i=1}^{N_1}S_i+\sum_{k=1}^{N_2}\tilde{S}_k,
  \label{Sloc}
\end{equation}
where
\begin{equation}
  S_i=\!{}-\frac{1}{g_o^2}\int d^4x\sqrt{-g}
  \frac{\Fc'(J_i)}{2}\left(g^{\mu\nu}\partial_\mu\phi_i\partial_\nu\phi_i
    +J_i\phi_i^2\right),
\end{equation}
\begin{equation}
  \begin{array}{l}
    \!\displaystyle\tilde{S}_k=\!\displaystyle\! {}-\frac{1}{g_o^2}\int
    d^4x\sqrt{-g}\left(g^{\mu\nu}\left(\frac{{
            \Fc''(\tilde{J}_k)}}{4}\left(\partial_\mu
          \tilde{\phi}_k\partial_\nu\varphi_k+\partial_\nu
          \tilde{\phi}_k\partial_\mu\varphi_k\right)+{}\right.\right.\\[2.7mm]
    \displaystyle + \left.\frac{
        \Fc'''(\tilde{J}_k)}{12}\partial_\mu\varphi_k\partial_\nu\varphi_k\right)+
    \left. \frac{\tilde{J}_k \Fc''(\tilde{J}_k)}{2}\tilde\phi_k\varphi_k
      +\left(\frac{{\tilde{J}_k \Fc'''(\tilde{J}_k)}}{12}+\frac{{
            \Fc''(\tilde{J}_k)}}{4}\right)\varphi_k^2\right), \label{Slocdr}
  \end{array}
\end{equation}
we can see that solutions of the Einstein equations and equations in
$\phi_k$, $\tilde{\phi}_k$ and $\varphi_k$, obtained from this action,
solve the initial  nonlocal equations (\ref{fr_equ}). Thus, we
obtain that special solutions to nonlocal equations can be found as
solutions to system of local (differential) equations. If $ \Fc(J)$ has
an infinity number of roots then one nonlocal model corresponds to
infinity number of different local models and the initial nonlocal
action (\ref{action_model2}) generates infinity number of local actions
(\ref{Sloc}).

We should prove that the way of localization is self-consistent.  To
construct local action (\ref{Sloc}) we assume that equations
(\ref{equphi}) are satisfied. Therefore, the method of localization is
correct only if these equations can be obtained from the local action
$S_{loc}$. The straightforward calculations show that the way of
localization is self-consistent because:
\begin{equation}
  \frac{\delta{S_{loc}}}{\delta \phi_i}=0 \, \Leftrightarrow \,
  \Box_g\phi_i=J_i\phi_i; \, \frac{\delta{S_{loc}}}{\delta
    \tilde{\phi}_k}=0 \, \Leftrightarrow \,
  \Box_g\varphi_k=\tilde{J}_k\varphi_k; \,
  \frac{\delta{S_{loc}}}{\delta \varphi_k}=0 \, \Leftrightarrow \,
  \Box_g\tilde{\phi}_k=\tilde{J}_k\tilde{\phi}_k+\varphi_k.
\end{equation}

In spite of the above-mention equations we obtain from $S_{loc}$ the
equations:
\begin{equation}
  G_{\mu\nu}=8\pi G_N\left(T_{\mu\nu}(\phi_0)-\Lambda g_{\mu\nu}\right),
\end{equation}
where $\phi_0$ is given by (\ref{phi0}) and $T_{\mu\nu}(\phi_0)$ can be
calculated by (\ref{Tmunugen}). So, we get such systems of
differential equations that any solutions of these systems are
particular solutions of the initial nonlocal equations (\ref{fr_equ}).

Let us consider functions $\Fc(J)$ with two and only two simple roots.
If $\Fc(J)$ has two real simple roots, then $\Fc'(J)>0$ at one root and
$\Fc'(J)<0$ at another root, so we get a quintom
model~\cite{Quinmodrev1}, in other words, local model with one standard
scalar field and one phantom scalar field. In the case of two complex
conjugated simple roots $J_j$ and $J_j^*$ one gets the following
action:
\begin{equation}
S_c=\!\int\!\! d^4x\frac{\sqrt{-g}}{2g_o^2}\left(
  \Fc'(J_j)\left(g^{\mu\nu}\partial_\mu\phi_j\partial_\nu\phi_j
    +J_j\phi_j^2\right)+{\Fc'}^*(J_j)\left(g^{\mu\nu}\partial_\mu\phi^*_j\partial_\nu\phi^*_j
    +J^*_j{\phi_i^*}^2\right)\right).
\end{equation}
We introduce real fields $\xi$ and $\eta$ such that $\phi_j=\xi+i\eta$,
\ $\phi_j^*=\xi-i\eta$, denote $d_r\equiv\Re e(\Fc'(J))$, \
$d_i\equiv\Im m(\Fc'(J))$, and obtain:
\begin{equation}
S_c=\int d^4x\frac{\sqrt{-g}}{2g_o^2}\Bigl(d_r
g^{\mu\nu}\left(\partial_\mu\xi\partial_\nu\xi-
\partial_\mu\eta\partial_\nu\eta\right)+
d_ig^{\mu\nu}(\partial_\mu\xi\partial_\nu\eta-\partial_\mu\eta\partial_\nu\xi)+V_1\Bigr),
\end{equation}
where $V_1$ is a potential term. In the case $d_i=0$ we get a quintom
model, in opposite case the kinetic term in $S_c$ has a nondiagonal
form. To diagonalize the kinetic term we make the transformation:
$\chi=\upsilon+\tilde{C}\sigma$, $\eta={}-\tilde{C}\upsilon+\sigma$,
where $\tilde{C}\equiv\left(d_r+\sqrt{d_r^2+d_i^2}\right)/d_i$, and get
a quintom model:
\begin{equation}
S_c=\int
d^4x\frac{\sqrt{-g}}{2g_o^2}\left(\frac{2\left(d_r^2+d_i^2\right)}{d_i^2}\left(d_r+\sqrt{d_r^2+d_i^2}\right)
\left(\partial_\mu\upsilon\partial_\nu\upsilon
-\partial_\mu\sigma\partial_\nu\sigma\right)+V_1\right).
\end{equation}

In the case of a real double root $\tilde{J}_k$ we express
$\tilde{\phi}_k$ and $\varphi_k$ in terms of new fields $\xi_k$ and
$\chi_k$:
\begin{eqnarray}
    \tilde{\phi}_k&=&\frac{1}{2\Fc''(\tilde{J}_k)}\left(\left(\Fc''(\tilde{J}_k)-\frac{2}{3}\Fc'''(\tilde{J}_k)\right)
    \xi_k-\left(\Fc''(\tilde{J}_k)+\frac{2}{3}\Fc'''(\tilde{J}_k)\right)\chi_k\right),
\quad     \varphi_k=\xi_k+\chi_k,\nonumber
\end{eqnarray}
we obtain the corresponding $\tilde{S}_k$ in the following form:
\begin{eqnarray}
\tilde{S}_k&=&\frac{{}-1}{2g_o^2}\!\int\!
d^4\!x\sqrt{-g}\left(g^{\mu\nu}\frac{\Fc''(\tilde{J}_k)}{4}(\pd_\mu
\xi_k\pd_\nu\xi_k-\pd_\nu
\chi_k\pd_\mu\chi_k)+\left[\frac{{\tilde{J}_k\Fc'''(\tilde{J}_k)}}{12}+\frac{{\Fc''(\tilde{J}_k)}}{4}\right]
(\xi_k+\chi_k)^2+\right.\nonumber\\
 &+&\left.\frac{\tilde{J}_k}{4}\left[(\Fc''(\tilde{J}_k)-\frac{2}{3}\Fc'''(\tilde{J}_k))
    \xi_k-(\Fc''(\tilde{J}_k)+\frac{2}{3}\Fc'''(\tilde{J}_k))\chi_k\right](\xi_k+\chi_k)\right).\nonumber
\end{eqnarray}
It is easy to see that each $\tilde{S}_k$ includes one phantom scalar
field and one standard scalar field. So, in the case of one double root
we obtain a quintom model. In the Minkowski space appearance of phantom
fields in models, when $\Fc(J)$ has a double root, has been obtained
in~\cite{PaisU}. So, we come to conclusion that both two simple roots
and one double root of $\Fc(J)$ generate quintom models.

The model with action (\ref{action_model2}) in the case $C_1\neq 0$ has
been considered in detail in~\cite{VernovSQS}. Here we present only the
obtained algorithm of localization for an arbitrary quadratic potential
$V(\phi)=C_2\phi^2+C_1\phi+C_0$:

\begin{itemize}
\item Change values of $f_0$ and $\Lambda$ such that the potential
  takes the form $V(\phi)=C_1\phi$.
\item Find roots of the function $ \Fc(J)$ and calculate orders of
 them. Select an finite number of simple and double roots.
\item Construct the corresponding local action. In the case $C_1=0$
  one should use formula (\ref{Sloc}). In the case $C_1\neq 0$ and
  $f_0\neq 0$ one should use (\ref{Sloc}) with the replacement of the
  scalar field $\phi$ by $\chi$ and the corresponding modification of
  the cosmological constant. In the case $C_1\neq 0$ and $f_0=0$ the
  local action is the sum of (\ref{Sloc}) and either
  \begin{equation}
    S_{\psi}={}-\frac{1}{2g_o^2}\int\! d^4x\sqrt{-g}\left(
      f_1g^{\mu\nu}\partial_\mu\psi\partial_\nu\psi+2C_1\psi+\frac{f_2C_1^2}{f_1^2}\right),
  \end{equation}
  in the case of simple root $J=0$, or
 \begin{eqnarray}
   S_{\tilde{\psi}}&=&{}-\!\int\! d^4x\frac{\sqrt{-g}}{2g_o^2}\left[
      g^{\mu\nu}\left(f_2(\partial_\mu\tilde{\psi}\partial_\nu\tau
        +\partial_\nu\tilde{\psi}\partial_\mu\tau)+f_3\partial_\mu\tau\partial_\nu\tau\right)
      +f_2\tau^2+2C_1\tilde{\psi}+\frac{f_3C_1}{2f_2}\tau\right]\nonumber
  \end{eqnarray}
  in the case of double root $J=0$. Note that in the case $C_1\neq 0$
  and $f_0=0$ the local action (\ref{Sloc}) has no term, which
  corresponds to the root $J=0$.
\item Vary the obtained local action and get a system of the Einstein
  equations and equations of motion. The obtained system is a finite
  order system of differential equations, \textit{i.e.} we get a local
  system. Seek solutions of the obtained local system.

\end{itemize}

\section{Conclusion}

The main result of this paper is the generalization of the algorithm of
localization on  the $f(R)$ gravity models with a nonlocal scalar
field. The algorithm of localization is proposed for an arbitrary
analytic function $\Fc(\Box_g)$, which has both simple and double
roots. We have proved that the same functions solve the initial
nonlocal Einstein equations and the obtained local Einstein equations.
We have found the corresponding local actions and proved the
self-consistence of our approach. In the case of two simple roots as
well as in the case of one double root we get a quintom
model~\cite{Quinmodrev1}. The algorithm of localization does not depend
on metric, so it can be used to find solutions for any metric.

The author wishes to express his thanks to I.~Ya.~Aref'eva for useful
and stimulating discussions. The research
  has been supported in part by RFBR grant 08-01-00798, grant of Russian
  Ministry of Education and Science NSh-4142.2010.2 and by Federal
  Agency for Science and Innovation under state contract
  02.740.11.0244.

\end{document}